\documentclass{article}
\usepackage{amsmath}

\begin{document}

\begin{center}
Geometrical Properties of Feynman Path Integrals

Timur F. Kamalov

Physics Department, Moscow State Open University

Korchagina, 22, Moscow, 107996, Russia E-mail:

ykamalov@rambler.ru, qubit@mail.ru
\end{center}

\textit{This model is one of the possible geometrical interpretations of
Quantum Mechanics where found to every image Path correspondence the
geodesic trajectory of classical test particles in the random geometry of
the stochastic fields background. We are finding to the imagined Feynman
Path a classical model of test particles as geodesic trajectory in the
curved space of Projected Hilbert space on Bloch's sphere.}



In this model we describe the experimental microobjects as test particles
with low mass and small size which get their quantum properties in the
random curved space. In our model there is corresponds Feynman's Path
Integrals to the geodesic trajectory of test particles in Projective Hilbert
space named Pro-Hilbert space. We show that it is possible find to Feynman's
description by imagined trajectories a classical model. We can to compare
imagined Feynman Path and trajectories of test particles with classical
action's function in random fields. In that case every trajectory in
Feynman's Path Integrals corresponds to influence of single field from
random fields background.

Recent years a very fascinating idea to put QM into geometric language
attracts the attention of many physicists~[1]. The starting point for such
an approach is the projective interpretation of the Hilbert space $\mathcal{H%
}$ as the space of rays. To illustrate the main idea it is convenient to
decompose the Hermitian inner product $\langle \cdot |\cdot \rangle $ in $%
\mathcal{H}$ into real and imaginary parts by putting for the two $L_{2}$%
--vectors $|\psi _{1}\rangle =u_{1}+\imath v_{1}$ and $|\psi _{2}\rangle
=u_{2}+\imath v_{2}$:

\begin{center}
\begin{equation}
\langle \psi _{1}|\psi _{2}\rangle =G\,(\psi _{1},\psi _{2})-\imath \Omega
\,(\psi _{1},\psi _{2}),
\end{equation}
\end{center}

where $G$ is a Riemannian inner product on $\mathcal{H}$ and $\Omega $ is a
symplectic form, that is

\begin{equation}
G\,(\psi _{1},\psi _{2})=(u_{1},u_{2})+(v_{1},v_{2});\quad \Omega \,(\psi
_{1},\psi _{2})=(v_{1},u_{2})-(u_{1},v_{2}),
\end{equation}

with $(\cdot ,\cdot )$ denoting standard $L_{2}$ inner product. The
symplectic form $\Omega $ revealed in (1) can acquire its dynamical content
if one uses the special stochastic representation of QM.

In our simulation of quantum trajectories we consider the behavior of
classical test particles which get the property analogue to
quantum-statistical property because the influence of random geometry. In
our case we consider a differential equations with random coefficients. For
example, the relative interval $\ell $\ between two test particles in
classical gravitational fields from random sources is described by deviation
equations of General Relativity Theory [2]. In our case we have the
stochastic deviation equation

\begin{center}
\begin{equation}
\frac{D^{2}}{D\tau ^{2}}\ell ^{i}(j,t)=R_{kmn}^{i}(j)\ell ^{m}(j,t)\frac{%
dx^{k}}{d\tau }\frac{dx^{n}}{d\tau }+f(j)
\end{equation}
\end{center}

Here $R_{kmn}^{i}(j)$- Riemann's tensor of gravitational fields from random
sources by different objects in Universe, $f(j)$- stochastic constant with
particular case $f(j)=0$). By the indexes $j=1,2,3,...$ we denoted the
number of the single field from the set of the objects in the Universe. Here
$i,k,m,n=0,1,2,3$. In our model random gravitational fields get stochastic
oscillations of relative intervals between two particles.

From (3) it follows that random gravitational fields get the oscillations,
which are described by the deviation equations. The solution of these
differentials equations is the exponent [3]. As it is easy to check it
satisfies the equations (3), \qquad

\begin{center}
\begin{equation}
\ell ^{i}(j)=\ell _{0}^{i}\exp k_{m}(j)x^{m}
\end{equation}
\end{center}

It is analogy to Feynman's kernel [4].

In 3-dimentional form we have $\ell ^{a}(j)=\ell _{0}^{a}\exp
(k_{b}(j)x^{b}+i\omega (j)t),a,b=1,2,3$ or,

\begin{center}
\begin{equation}
\ell ^{i}(j)=\ell _{0}^{i}\exp \frac{1}{\hbar }S(j)
\end{equation}
\end{center}

Where $S(j)=p^{m}x_{m}$ is action function.

In particular case of one-dimension oscillations of the deviation equations
we have the equations of the oscillations of two particles

\begin{center}
\begin{equation}
\overset{\cdot \cdot }{\ell }^{1}(j,t)+c^{2}R_{010}^{1}(j)\ell
^{1}(j,t)=0,\omega (j)=c\sqrt{R_{010}^{1}(j)}
\end{equation}
\end{center}

It is easy to check that the exponent $\ell ^{1}(j)=\ell _{0}^{1}\exp
(k_{a}(j)x^{a}+i\omega (j)t)$ satisfies the equation (4) with the indexes $%
a=1,2,3$. The quantitative $\ell (j)$ with the random phase $\Phi
(j,t)=i\omega (j)t$\ corresponds to each gravitational field with the index $%
j$ and Riemann's tensor $R_{kmn}^{i}(j)$. In the common case the random
phase is

\begin{center}
\begin{equation}
\Phi (j)=\int i\omega (j)dt=\frac{i}{\hbar }S(j)
\end{equation}
\end{center}

In the common case for having the result we must sum on the gravitational
fields

\begin{center}
\begin{equation}
S_{0}=\sum S(j)=p_{i}x^{i}
\end{equation}
\end{center}

From General Relativity Theory the action function in the vacuum with the
metric $g_{ik}$ of gravitational fields is

\begin{center}
\begin{equation}
S(j)=-\frac{1}{16\pi G}\int (dx)\sqrt{-g(j)R(j)}
\end{equation}
\end{center}

Let's compare (4) with the function

\begin{center}
\begin{equation}
\varphi \lbrack \ell (j,t)]=C\exp \{\frac{i}{\hbar }S[\ell (j,t)]\}
\end{equation}
\end{center}

And let's substitute this function to (8), (9). Then Feynman's kernel for
the movement of our object from the point a to b with the quantitative (5) is

\begin{center}
\begin{equation}
K(b,a)=\sum \varphi \lbrack \ell (j,t)]
\end{equation}
\end{center}

In our case every phase of the trajectory is proportional to the action
function

\begin{equation}
\varphi \lbrack \ell (j,t)]=C\exp \{\frac{i}{S_{0}}S[\ell (j,t)]\}
\end{equation}

For the interval $\Delta \ell $\ we have the probability distribution of the
particle

\begin{equation}
\Delta P=\frac{1}{\sigma _{\ell \sqrt{2\pi }}}\exp (-\frac{\Delta \ell ^{2}}{%
2\sigma _{\ell }})
\end{equation}

Here $\sigma _{\ell }$\ is a constant of the distribution function.

For this distribution function we can write

1.\qquad If the interval of the particle distribution $\Delta \ell
^{2}=g_{ik}\Delta x^{i}\Delta x^{k}$\ in Pseudo-Riemann's space is small
i.e. extends to zero then the probability of the observing the particle in
this interval extends to unit.

2.\qquad If the interval $\Delta \ell $\ in the Pseudo-Riemann's space
extends to the infinitive then the probability to observe this particle in
this interval extends to the zero.

3.\qquad The probability to observe the particle in intervals $\Delta \ell
(x_{2},x_{1})+\Delta \ell (x_{3},x_{2})\geq \Delta \ell (x_{3},x_{1})$\ is $%
P_{21}+P_{32}\leq P_{31}$.

4.\qquad If the interval extends to the zero then probability to observe the
particle extend to the unit.

Here and later we have deal with the similar-time intervals or the
similar-space intervals.

5. If the interval extends to the infinite quantities than probability is
zero.

For the probability interval to observe the particle in the velocity
interval $\Delta u^{i}$\ is

\begin{equation}
\Delta P=\frac{1}{\sigma \sqrt{2\pi }}\exp (-\frac{(\Delta u^{i})^{2}}{%
2\sigma ^{2}})
\end{equation}

or

\begin{equation}
\Delta P=\frac{1}{\sigma \sqrt{2\pi }}\exp (-\frac{S}{S_{0}})
\end{equation}

where $\sigma $- is a constant for this distribution, $m$- is the mass of
this particle, $W$- is the energy of the oscillation for the particle with
its have in random fields. For the single particle root-mean-square
deviation $(\Delta u^{i})^{2}$\ of the velocity is connected to the action
function $S$. We can think that the dispersion is a characteristic of the
geometrical background and then we have

\begin{equation}
\Delta P=\frac{1}{\sigma \sqrt{2\pi }}\exp (-\frac{W}{2\sigma ^{2}})
\end{equation}

where we denotes $a^{2}=\frac{1}{\sigma \sqrt{2\pi }}$\ probability
amplitude, $S_{0}=\frac{m\sigma ^{2}}{2}$- is the action function for the
particle with the mass $m$. Than Schredinger's wave-function is corresponds
to the geometrical function

\begin{equation}
\psi =a\exp (i\frac{S}{2S_{0}})
\end{equation}

which mean the angle correspondences to the area S on the spherical surface
of the phase space $p$\ and $x$.

In 3-dimentional space we can write the classical Hamilton's equation

\begin{equation}
\frac{\partial S}{\partial t}+\frac{1}{2m}(\nabla S)^{2}+U=0
\end{equation}

with the continuous equation for the probability density $\psi ^{2}=a^{2}$,

\begin{equation}
\frac{\partial a^{2}}{\partial t}+div(a^{2}\frac{\nabla S}{m})=0
\end{equation}

Than we have the analogue of the Schredinger's equation for the geometrical
function (17) defined by (9)

\begin{equation}
i2S_{0}\frac{\partial \psi }{\partial t}=-\frac{4S_{0}^{2}}{2m}\Delta \psi
+U(x,y,z)\psi
\end{equation}

When we sum the classical path integrals for the geodesic trajectories from
random curved space considered as geometrical interpretation of the image
Feynman's trajectories. In this case we used test particles with the small
mass and size for the non-perturbation observation.

Author thanks to M.L. Filchenkov and K.S.Kabisov for the helpful discussions.

\begin{center}
\textit{References}
\end{center}

[1] \textit{D.~C.~ Brody and L.~P.~ Hughston}. Geometric Quantum Mechanics
// \textit{J. of Geometry and Physics}. 2001, vol. \textbf{38}. Pp. 19--53.

[2] \textit{S. Weinberg. }Gravitation and cosmology, J. Wiley and Sons,
Inc., N.Y., 1972.

[3] \textit{C. W. Misner, K.S. Torn, J.A. Wheeler.} Gravitation, W.H.
Freeman and Company. San Francisco. 1973.

[4] \textit{R. Feynman, A. Hibs.} Quantum Mechanic and Path Integrals, N.Y.:
McGraw-Hill Book Company, 1965.

\end{document}